\documentclass[twocolumn,aps,prl,showpacs,superscriptaddress,twocolumn]{revtex4-1}
\usepackage{lmodern}

\usepackage[latin9]{inputenc}
\usepackage{amsmath}
\usepackage{amssymb}
\usepackage{graphicx}
\usepackage{epstopdf}
\usepackage{color}
\usepackage{enumerate}
\usepackage{ulem}
\usepackage{natbib}
\usepackage{mathtools}
\usepackage{caption}
\usepackage{soul,color}
\definecolor{orange}{RGB}{255,125,0}

\begin{document}
	
	\bibliographystyle{apsrev}
	
	\title{Instability of insulators near quantum phase transitions}

	\author{A. Doron}
	\email{adam.doron@weizmann.ac.il; Corresponding author}
	\affiliation{Department of Condensed Matter Physics, The Weizmann Institute of Science, Rehovot 76100, Israel.}
	
	\author{I. Tamir}
	\affiliation{Department of Condensed Matter Physics, The Weizmann Institute of Science, Rehovot 76100, Israel.}
	
	\author{T. Levinson}
	\affiliation{Department of Condensed Matter Physics, The Weizmann Institute of Science, Rehovot 76100, Israel.}
	
	\author{M. Ovadia}
	\affiliation{Department of Condensed Matter Physics, The Weizmann Institute of Science, Rehovot 76100, Israel.}
	\affiliation{Present Address: Department of Physics, Harvard University, Cambridge, Massachusetts 02138, United States.}
	
	\author{B. Sac\'ep\'e}
		\affiliation{Univ. Grenoble Alpes, CNRS, Grenoble INP, Institut N\'{e}el, 38000 Grenoble, France}	
	
	\author{D. Shahar}
	\affiliation{Department of Condensed Matter Physics, The Weizmann Institute of Science, Rehovot 76100, Israel.}

	\begin{abstract}
	Thin films of Amorphous indium oxide undergo a magnetic field driven superconducting to insulator quantum phase transition. In the insulating phase, the current-voltage characteristics show large current discontinuities due to overheating of electrons. We show that the onset voltage for the discontinuities vanishes as we approach the quantum critical point.
	As a result the insulating phase becomes unstable with respect to any applied voltage making it, at least experimentally, immeasurable.
	We emphasize that unlike previous reports of the absence of linear response near quantum phase transitions, in our system, the departure from equilibrium is discontinuous. Because the conditions for these discontinuities are satisfied in most insulators at low temperatures, and due to the decay of all characteristic energy scales near quantum phase transitions, we believe that this instability is general and should occur in various systems while approaching their quantum critical point. Accounting for this instability is crucial for determining the critical behavior of systems near the transition. 
	\end{abstract}
	
	\maketitle

The superconducting insulator transition (SIT)  \cite{goldmanpt51,physupekhi}, observed in highly disordered superconductors, is a quantum phase transition (QPT) \cite{sondhirmp} driven by varying the magnetic field ($B$) \cite{HebardPrl,kapitulnikprl74,BaturinaJETP}, disorder \cite{Shaharprb}, film thickness \cite{haviprl62} or charge density \cite{goldmanprl94}.

\begin{figure} [h!]
		\includegraphics [width=8.5 cm] {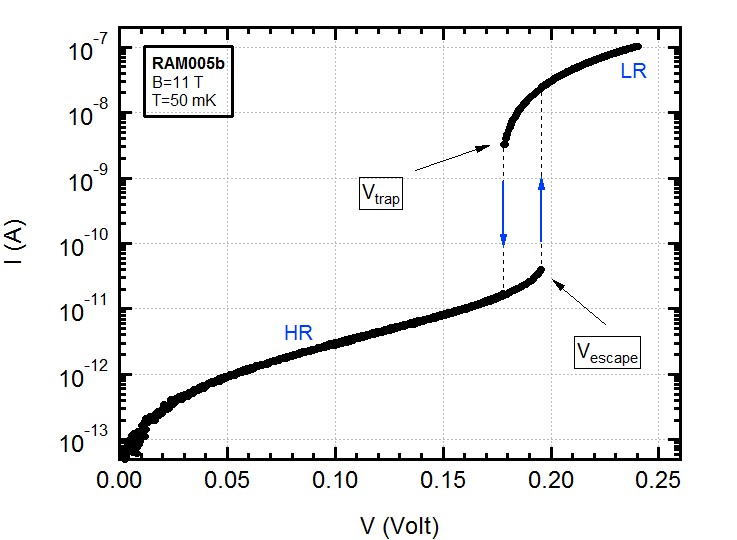}
	
		\caption{{\bf Discontinuities in the $I-V$'s.} 
			$I$ (log scale) vs. $V$ measured at $T=50$ mK and $B=11$ T (in the insulating phase). The measured data points are marked by full circles, the dashed line connecting the data points is a guide for the eye. The LR$\to$HR (HR$\to$LR) transition is marked by a blue arrow pointing down (up).
			At $V=0$ the sample is in the HR state. By increasing $V$ the $I-V$'s exhibit a discontinuity at $V_{escape}=0.195$ Volts where $I$ jumps by 3 orders of magnitude.
			Decreasing $V$ results in a hysteresis where $I$ drops back to the HR state at $V_{trap}=0.178$ V.
		}			
		\label{Figure1}
	\end{figure}

In the $B$-driven SIT, beyond the critical $B$ ($B_{c}$), Cooper-pairs persist and become spatially localized \cite{FeigAnnals,YonatanNat,GantmakherJETP,murthyprl,vallesprl103,BenjaminNat,sacepe2015high}, leading to a strongly insulating state \cite{paalanenprl69,GantmakherJETP,murthyprl2004,vallesprl103,benjaminprl101,kapitulnikprl74}. 
In this insulating state, at low $T$ ($T<200$ mK), the current-voltage characteristics ($I-V$'s) exhibit large $I$-discontinuities ($\Delta I$) \cite{murthyprl} (figure \ref{Figure1}). 
The $I-V$'s can be separated into two regions, a high resistance (HR) state at low $V$ and a low resistance (LR) state at high $V$.
We denote the threshold $V$ where the HR$\to$LR (LR$\to$HR) occurs as $V_{escape}$ ($V_{trap}$)

Recently Altshuler \textit{et al.} showed that the $\Delta I$'s could be explained by a thermal bi-stability where the electrons can thermally decouple from the phonon bath resulting in a well defined electron $T$ ($T_{el}$) \cite{borisprl}.
The central assumption of this model is that deviations from a linear $I-V$ result from an increase in $T_{el}$.
The steady state $T_{el}$ is determined by the heat-balance between the experimentally applied Joule-heating ($I\cdot V$) and cooling via the phonons.
This non-equilibrium state is analyzed by solving the heat-balance equation
	\begin{equation}
			\frac{V^{2}}{R(T_{el})}=\Gamma \Omega (T_{el}^{\beta}-T_{ph}^{\beta})
			\label{eHB}
	\end{equation}
where $\Omega$ is the volume of the sample, $\Gamma$ is the electron-phonon coupling-strength, $\beta$ is an exponent that determines the power-law decay of the electron-phonon coupling as $T \to 0$ and $R(T_{el})=R_{0}exp((\Delta /T_{el})^{\gamma})$, typical of insulators ($\Delta$ is the insulator's activation energy and $\gamma$ is typically $\le 1$).

The central result of this model is that, below a critical phonon $T$ ($T_{ph}^{cr}$), equation (\ref{eHB}) has two stable solutions for $T_{el}$. The $\Delta I$'s are a result of a change in $R$ that occurs when the electrons abruptly switch between the low $T_{el}$ solution, where the HR state exists, and the high $T_{el}$ solution. 
This electron-heating approach gained support from several experiments \cite{maozprl, KalokArxiv,levinson2016direct}.

% % % % % % % % % % %  % 
%					   %
% End of introduction  % 
%					   %
% % % % % % % % % % %  % 

At first sight, the far-from-equilibrium $\Delta I$'s and its underlying electron-phonon decoupling appear to be not relevant to the study of the equilibrium phases, and of the SIT itself. The main conclusion from the results presented in this Letter is that near the quantum critical point (QCP) of the SIT this is not the case. 
By systematically followed the $B$-- evolution of $\Delta I$ we show that $V_{escape}$ vanishes as $B\to B_{c}$. Consequently, close enough to $B_{c}$, the finite $V$ required for transport measurements will inevitably exceed $V_{escape}$, driving the system out of equilibrium with a discontinuous transition to the high $T_{el}$ state. % making the HR state unaccessible experimentally. 
The significance of the discontinuous departure from equilibrium is that, not only the equilibrium state cannot be measured but near the QCP, where like all other energy scales $V_{escape}\to0$, it is not experimentally possible to extrapolate the equilibrium properties from the measurable $R$.

Our data were obtained by measuring thin-films of amorphous indium-oxide. The films were deposited by e-gun evaporation of In$_2$O$_3$ onto a SiO$_2$ substrate in an  O$_2$ rich environment. Both samples have a Hall-bar geometry, their lengths and widths are $2 \times 0.5$ mm$^2 $ (sample RAM005b) and $10 \times 5 \mu$m$^2$ (BT1c) and their thickness is $30$ nm (see section S2 \cite{Supplemental} for a discussion about the contact). The data presented were measured in a two-probe dc configuration, which agrees with our 4-terminal measurements in their overlapping regime of applicability.

\begin{figure} [h!]
		\includegraphics [width=8.5 cm] {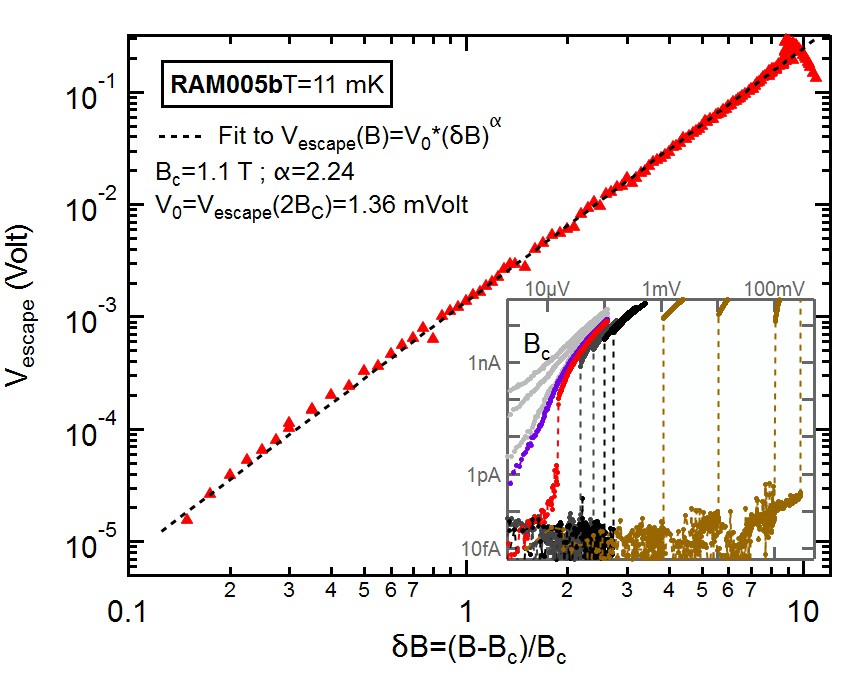}
					
		\caption{{\bf Magnetic-field evolution of $\boldsymbol V_{escape}$.} 
			$V_{escape}$ vs $\delta B \equiv \frac{B-B_{c}}{B_{c}}$ (log-log scale) at $T=11$ mK. 
			The data are presented as red triangles while a dashed black-line marks a power-law fit. 
			Inset: $I$ vs $V$ (log-log scale) at 11 mK measured at $B= 10, 8, 3.5, 2$ T (brown $I-V$'s), $B=1.45, 1.4, 1.35, 1.3$ T (black),  $B=1.25$ T (red), $B=1.225$ T (purple) and $B=1.2, 1.15, 1.1$ T (gray).
				}			
		\label{Figure2}
	\end{figure}

The main results of our work are summarized in figure \ref{Figure2}, were we display $V_{escape}$ vs. $\delta B \equiv \frac{B-B_{c}}{B_{c}}$ at our base $T=11$ mK.
On this log-log plot $V_{escape}$ follows a power-law that spans almost two decades in $\delta B$ and four decades in $V_{escape}$, indicating that $V_{escape}$ vanishes upon approaching $B_{c}$,	
	\begin{equation}
	V_{escape}(B) \propto (B-B_{c})^\alpha
	\label{ePower-Law}
	\end{equation}
where $\alpha=2.24$ is extracted using a power-law fit ($\alpha$ appears to be non-universal: it is sample and $T$ dependent).

Equation \ref{ePower-Law} reflects the inherent experimental difficulty one is faced when measuring equilibrium properties near $B_{c}$.
While conducting transport measurements, it is essential to apply a finite $V$ across the sample. 
This applied $V$ must exceed the noise present during the experiment (either instrumental or inherent such as Johnson-Nyquist noise) and must also be large enough to induce a measurable $I$ response from the sample (typically $V>\mu$V). 
The vanishing of $V_{escape}$ suggests a loosing cause: whatever small $V$ is, there will always be a $B$ range, close to the SIT, where it will exceed $V_{escape}$ and drive the system out of equilibrium. 

In the inset of figure \ref{Figure2} we display $I-V$'s from which the data of the main figure were extracted.
	Close to $B_{c}$ (black color) $V_{escape}$ decreases down to $B=1.25$ T (red), which is the lowest $B$ where a discontinuity was observed (at $V_{escape}$=15$\mu$V).
	According to the power-law fit of equation (\ref{ePower-Law}), $V_{escape}(B=1.225$ T$)\sim9\mu$V, but, at $B=1.225$ T (purple) the data already appears continuous. 
	A possible reason for this is that for this $B$ range, the measurement $T=11$ mK might become larger than  $T_{ph}^{cr}$. As we discuss in section S1 \cite{Supplemental} this is unlikely.
	A more probable explanation is that the integrated voltage-noise surpasses $V_{escape}$, the $I-V$'s will appear continuous and measure only the LR state.

It is known that systems exhibit a non-linear response near QPT's \cite{sondhirmp,green2005nonlinear,hogan2008universal,dalidovich2004nonlinear} where, at $T=0$, any finite $V$ will drive them out of equilibrium. 
The pivotal difference reported in this letter is that, not only our system has no linear response, but it departs from equilibrium in a discontinuous fashion. 
In the discussion section we present several consequences of this discontinuous response.

\begin{figure} [h!]
		\centering
		\includegraphics [width=8.5 cm] {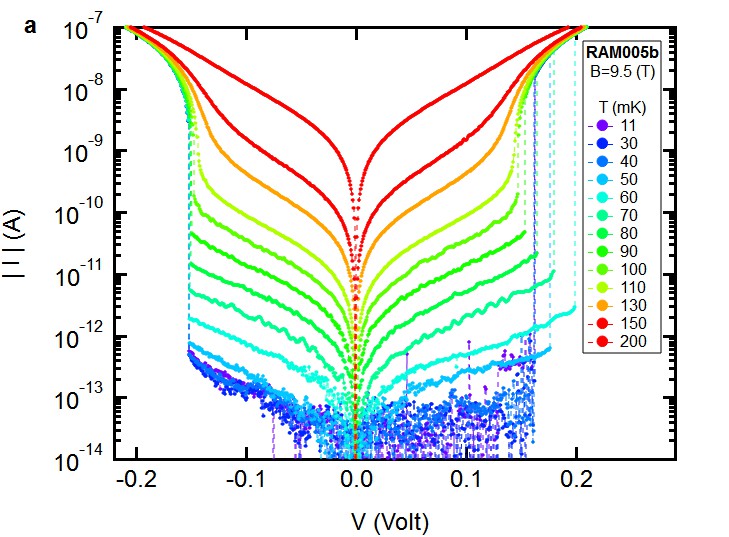}
		\includegraphics [width=8.5 cm] {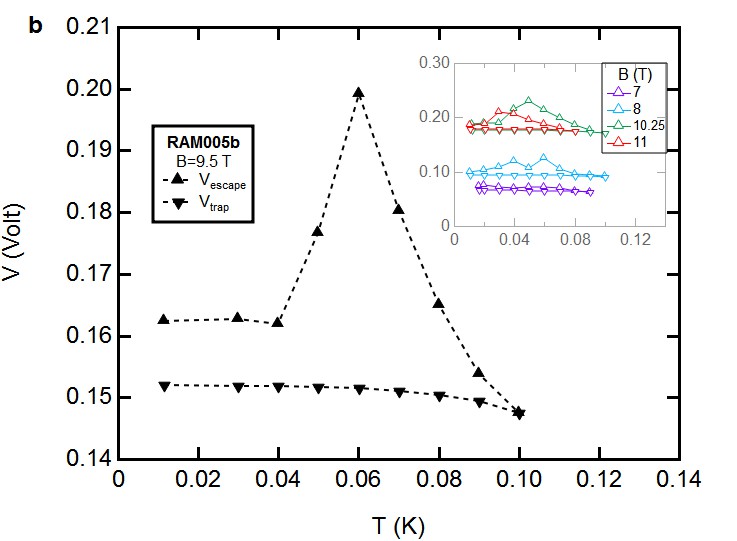}				
		\caption{{\bf Saturation of $\boldsymbol{ V_{escape}}$ as $\boldsymbol{T_{ph}\to 0}$.} 
			(a) $|I|$ (log scale) vs. $V$ measured at $B=9.5$ T. The color coding describe different $T_{ph}$ isotherms ranging from $11$ mK (purple) to $200$ mK (red).  At $T_{ph}=100$ mK, the $I-V$'s become discontinuous with $V_{escape}\sim 150$ mV. While cooling, $V_{escape}$ initially increases up to $V_{escape}=200$ mV at $T_{ph}=60$ mK. At lower $T_{ph}$, $V_{escape}$ drops and saturates at a finite $V$.
			(b) $V_{escape}$ and $V_{trap}$ vs. $T_{ph}$ at $B=9.5$ T. $V_{escape}$ and $V_{trap}$ are marked by up and down pointing triangles respectively. At low $T$'s, $V_{escape}$ saturates at a value which is comparable to $V_{trap}$. \textit{Inset:} $V_{escape}$ and $V_{trap}$ vs. $T_{ph}$ at different $B$'s.% All $B$'s show qualitatively similar behavior, $V_{trap}$ is almost $T_{ph}$ independent while $V_{escape}$ initially increases and eventually saturates at $V_{escape}\gtrsim V_{trap}$.
		}			
		\label{Figure3}
	\end{figure}

	Because, strictly speaking, the QPT takes place at $T=0$ it is worthwhile to examine the $T$ evolution of $V_{escape}$. If $V_{escape}$ increases sufficiently as $T \to 0$, at low enough $T$'s, the HR state might span a large $V$-interval and become measurable.
	In Figure \ref{Figure3}a we display the $I-V$'s measured at $B=9.5$ T at different $T$'s. Below $T=100$ mK, the $I-V$'s become discontinuous. As predicted \cite{borisprl}, $V_{trap}$ (the $\Delta I$ at $V<0$) is nearly independent of the phonon $T$ ($T_{ph}$). $V_{escape}$, on the other hand, initially increases as $T_{ph}$ is reduced down to $T_{ph}=60$ mK (green to light blue) as expected. As $T_{ph}$ is further lowered (dark blue), the trend changes and $V_{escape}$ begins to decrease.
	In figure \ref{Figure3}b we display $V_{escape}$ and $V_{trap}$ (up and down pointing triangles respectively) vs $T$.
	It appears that $\lim\limits_{T_{ph}\to 0}V_{escape}(T_{ph})\sim V_{trap}$.
	In the inset of figure \ref{Figure3}b we display $V_{escape}$ and $V_{trap}$ vs $T$ at various $B$'s. For all measured $B$'s, $V_{escape}$ follows a similar pattern. 
	These results suggest that lowering $T$ will not increase $V_{escape}$ and make the HR state measurable.
	In the discussion we show that this $T$ dependence of $V_{escape}$ can be explained by considering some inhomogeneity in the system.

\begin{figure} [h!]
		\includegraphics [width=8.5 cm] {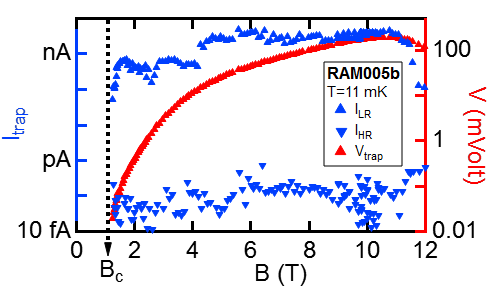}			
		\caption{{\bf $\boldsymbol B$ dependence of the $\boldsymbol \Delta I$'s.} 
			$I_{trap}\equiv I(V_{trap})$ and $V_{trap}$ vs. $B$.
			The blue triangles correspond to the left (blue) axis and mark $I$ on both sides of $V_{trap}$ where the upwards (downwards) pointing triangles correspond to the LR (HR) side of the jump.
			The red triangles correspond to the right (red) axis and mark $V_{trap}$.
			The vertical dashed black-line marks $B_{c}=1.1$ T.
		}			
		\label{Fig_R_HR_R_LR}
	\end{figure}

	An important question is how does the magnitude of the $\Delta I$'s evolve while approaching the QCP.
	If the $\Delta I$'s vanishes sufficiently fast, the transition from the LR to the HR states becomes practically continuous in the sense that one could extrapolate equilibrium, HR, properties from the LR state. 
	The $B$-dependence of the $\Delta I$'s at $T=11$ mK is displayed in figure \ref{Fig_R_HR_R_LR} where we focus on the trapping side where the LR$\to$HR transition occurs.
	The blue triangles correspond to the left (blue) axis and represent $I$ on both sides of $V_{trap}$ where upwards pointing triangles correspond to the last measured $I$ in the LR state before the jump ($I_{LR}$) and downwards pointing triangles stand for $I_{HR}$, the first measured $I$ in the HR state (for most $B$'s $I_{HR}$ was in the noise level). 
	The red triangles mark $V_{trap}$ and correspond to the right (red) axis.
	While $V_{trap}$ vanishes rapidly over a vast $B$ range, the magnitude of the $\Delta I$'s does not seem to vary significantly. 
	This observation has a great impact on the reliability of $I$-bias transport measurements as we will argue in the discussion.

\textit{Discussion} The data presented raise several points.
We demonstrated above the instability of the insulating state near the SIT. It is interesting to consider the relevance of our findings to other systems exhibiting a QPT involving insulators such as the metal-insulator and the quantum Hall transitions. According to reference \cite{borisprl}, $V_{trap}\propto \Delta ^{\beta/2}$, guaranteeing its vanishing at the SIT. Because all relevant energy scales, including $\Delta$ of the insulators, must vanish at all QCP, we expect similar inherent difficulties to arise in all QPT's involving insulators. (provided that $\beta > 0$). In sections S6 \cite{Supplemental} we display preliminary results we obtained by measuring the $I-V$'s of a silicon MOSFET sample in the insulating phase near the metal-insulator transition. These results strongly support our claim as, similarly to the SIT data, they also show discontinuous $I-V$'s with $V_{escape}$ vanishing while approaching the QPT.

\begin{figure} [h!]

			\includegraphics [width=8.5 cm] {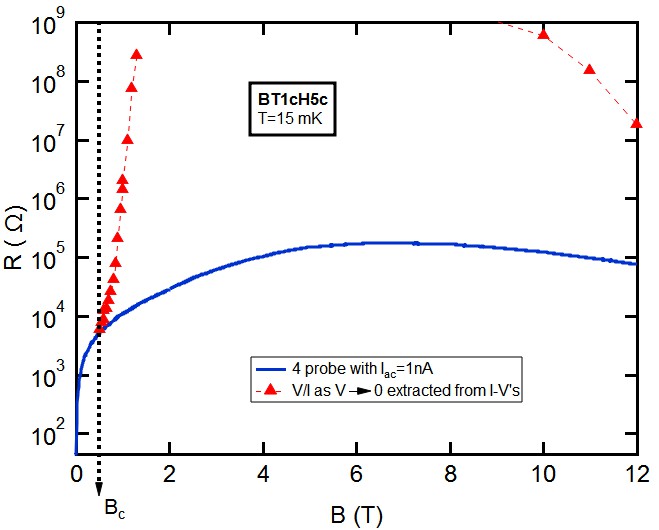}						
			\caption{{\bf Unavoidable "transport catastrophe".} 
				$R$ (log scale) vs. $B$ of sample BT1c at $T=15$ mK. The red triangles were extracted from the $I-V$'s in the limit $V \to 0$. The blue line was measured using a standard 4-probe AC measurement with $I_{rms}=1$ nA. 
			}			
			\label{Figure5}
		\end{figure}

So far we have stated the difficulties in $V$-biased measurements and showed that the close-to-equilibrium HR state can only be probed far from $B_{c}$. 
It turns out that there are similarly severe implications regarding $I$-biased, 4-terminal measurements. 
These are vividly illustrated in figure \ref{Figure5} where we plot data obtained from sample BT1c at $T=15$ mK. The red triangles are $R$ values extracted from the $V=0$ limit of two-probe DC $I-V$'s and represent our best estimation of the true Ohmic $R$ with the caveat that, close to $B_{c}$, we probably probe the LR state. 
In contrast, the blue line is a result of a standard, low frequency ($<$10 Hz), AC 4-probe measurement with $I_{rms}=1$ nA \cite{Footnote1na}. 
A pronounced discrepancy reaching several orders of magnitude exists between the two types of measurement for $B>B_c$. 
The roots of this discrepancy can be traced to figure \ref{Fig_R_HR_R_LR}. 
The 1 nA used in the 4-terminal measurement in figure \ref{Figure5} falls in the unstable regime of the $I-V$'s and far exceeds the maximum $I$'s allowed to observe the HR state. For sample BT1c $I=1$ nA is in the LR state for all $B>B_{c}$ (see figure S2a of \cite{Supplemental} for the $I-V$'s).

The overwhelming discrepancy in $R$'s between the two types of measurement displayed in figure \ref{Figure5} emphasizes that, for insulators at low $T$'s, transport data without studying the $I-V$'s are unreliable.
In sections S3 \cite{Supplemental} we show that the 4-probe R can be extracted from the LR state, in S4 \cite{Supplemental} we show that this discrepancy also occurs in larger samples, in S5 we discuss the 2 and 4 probe measurement configurations \cite{Supplemental}.

A notable feature in Figure \ref{Figure3}b is the non-monotonic $T$-dependence of $V_{escape}$, which initially increases on lowering $T$, reaches a maximum at $T=60$ mK, then decreases and eventually saturates at low $T$. 
 Altshuler \textit{et al.} noted that the $\Delta I$'s occur within a bi-stability $V$-interval ($\Delta V$), $V_{trap}^{min}<V<V_{escape}^{max}$, where $V_{trap}^{min}$ and $V_{escape}^{max}$ are the lower and upper bounds of $V_{trap}$ and $V_{escape}$ satisfying the parametric dependences: 
		
		\begin{equation}
		\begin{multlined}
		\quad V_{trap}^{min}\propto \Delta^{\frac{\beta}{2}}\\V_{escape}^{max}\propto \Delta^{-\frac{\gamma}{2}}T_{ph}^{\frac{\beta+\gamma}{2}}e^{\frac{1}{2}(\frac{\Delta}{T_{ph}})^{\gamma}}
		\label{eVth}
		\end{multlined}
		\end{equation}

Recalling that $\lim\limits_{B\to B_{c}}\Delta=0$ these equations imply that, as $T_{ph} \to 0$ and $B \to B_{c}$, $V_{trap}^{min}\to 0$ and $V_{escape}^{max}\to \infty$. Our data show that, for $T< 60$ mK, both $V_{trap}$ and $V_{escape}$ occur prematurely i.e. at the low limit of the $\Delta V$.
This can be understood by considering a mapping that we established between our non-equilibrium $I$-discontinuities and first-order phase transitions in a Van der Waals liquid \cite{DoronPRL}, where the actual transition occurs not at the limit of stability but according to the Maxwell area-rule. Only in near-ideal samples where no nucleation centers  are found, it is possible to observe a supercooled liquid close to the limit of stability \cite{garside1982nucleation}. If $V_{escape}$ is governed by a Maxwell-law equivalent, it is expected to increase upon cooling \cite{FootnoteVDW}. This is the case initially but, for $T< 60$ mK (Figure \ref{Figure3}b), an opposite trend develops. 

These premature jumps can be explained by inhomogeneity in the samples\cite{KravtsovPrivate} (indications for non-structural inhomogeneity were reported in similar systems \cite{BenjaminNat,kowal2008scale,kowal1994disorder,ioffe2012quantum,benjaminprl101}). At low $T$'s, there is a competition between inefficient cooling via phonons and the small Joule-heating (due to the sample's large $R$). While cooling is probably less sensitive to imperfections, heating may be affected near the impurities giving rise to "hot spots" that act similarly to nucleation centers in the Van der Waals liquid.

We would like to note an earlier work \cite{parendo2006hot} that remarks on the possibility that several observations near the SIT are due to electron heating, and shows that one cannot consider electric field and $T$ scaling independently. They did not consider discontinuous responses. 

In summary, our findings questions the stability of an equilibrium, insulating, state bordering the SIT.
Our main result is that $V_{escape}$, above which only the LR state persists, vanishes as we approach the QCP ($T\to 0$ and $B\to B_{c}$). 
This seemingly innocuous behavior has far-reaching consequences on transport:
Because transport measurements require a finite $V$ in order to probe the sample, only the LR state can be accessed. 
In addition, any $V$ noise will already heat the electrons and drive the sample to the LR state. 
A central question that remains is whether, theoretically, there is a $V$ range where the HR state is stable.

\begin{acknowledgments}
We are grateful to S. Kravchenko for providing a Si-MOSFET sample and to V. Kravtsov and I. Aleiner for fruitful discussions. This work was supported by the Israeli Science Foundation grant number 556/17.
\end{acknowledgments}

%\bibliography{Instability_of_insulators_near_quantum_phase_transitions}

\begin{thebibliography}{36}
	\expandafter\ifx\csname natexlab\endcsname\relax\def\natexlab#1{#1}\fi
	\expandafter\ifx\csname bibnamefont\endcsname\relax
	\def\bibnamefont#1{#1}\fi
	\expandafter\ifx\csname bibfnamefont\endcsname\relax
	\def\bibfnamefont#1{#1}\fi
	\expandafter\ifx\csname citenamefont\endcsname\relax
	\def\citenamefont#1{#1}\fi
	\expandafter\ifx\csname url\endcsname\relax
	\def\url#1{\texttt{#1}}\fi
	\expandafter\ifx\csname urlprefix\endcsname\relax\def\urlprefix{URL }\fi
	\providecommand{\bibinfo}[2]{#2}
	\providecommand{\eprint}[2][]{\url{#2}}
	
\bibitem[{\citenamefont{Goldman and Markovic}(1998)}]{goldmanpt51}
	\bibinfo{author}{\bibfnamefont{A.~M.} \bibnamefont{Goldman}} \bibnamefont{and}
	\bibinfo{author}{\bibfnamefont{N.}~\bibnamefont{Markovic}},
	\bibinfo{journal}{Phys. Today} \textbf{\bibinfo{volume}{51}},
	\bibinfo{pages}{39} (\bibinfo{year}{1998}).
	
	\bibitem[{\citenamefont{Gantmakher and Dolgopolov}(2010)}]{physupekhi}
	\bibinfo{author}{\bibfnamefont{V.~F.} \bibnamefont{Gantmakher}}
	\bibnamefont{and} \bibinfo{author}{\bibfnamefont{V.~T.}
		\bibnamefont{Dolgopolov}}, \bibinfo{journal}{Phys.-Usp.}
	\textbf{\bibinfo{volume}{53}}, \bibinfo{pages}{1} (\bibinfo{year}{2010}).
	
	\bibitem[{\citenamefont{Sondhi et~al.}(1997)\citenamefont{Sondhi, Girvin,
			Carini, and Shahar}}]{sondhirmp}
	\bibinfo{author}{\bibfnamefont{S.~L.} \bibnamefont{Sondhi}},
	\bibinfo{author}{\bibfnamefont{S.~M.} \bibnamefont{Girvin}},
	\bibinfo{author}{\bibfnamefont{J.~P.} \bibnamefont{Carini}},
	\bibnamefont{and} \bibinfo{author}{\bibfnamefont{D.}~\bibnamefont{Shahar}},
	\bibinfo{journal}{Rev. Mod. Phys.} \textbf{\bibinfo{volume}{69}},
	\bibinfo{pages}{315} (\bibinfo{year}{1997}).
	
	\bibitem[{\citenamefont{Hebard and Paalanen}(1990)}]{HebardPrl}
	\bibinfo{author}{\bibfnamefont{A.~F.} \bibnamefont{Hebard}} \bibnamefont{and}
	\bibinfo{author}{\bibfnamefont{M.~A.} \bibnamefont{Paalanen}},
	\bibinfo{journal}{Phys. Rev. Lett.} \textbf{\bibinfo{volume}{65}},
	\bibinfo{pages}{927} (\bibinfo{year}{1990}).
	
	\bibitem[{\citenamefont{Yazdani and Kapitulnik}(1995)}]{kapitulnikprl74}
	\bibinfo{author}{\bibfnamefont{A.}~\bibnamefont{Yazdani}} \bibnamefont{and}
	\bibinfo{author}{\bibfnamefont{A.}~\bibnamefont{Kapitulnik}},
	\bibinfo{journal}{Phys. Rev. Lett.} \textbf{\bibinfo{volume}{74}},
	\bibinfo{pages}{3037} (\bibinfo{year}{1995}).
	
	\bibitem[{\citenamefont{Baturina et~al.}(2004)\citenamefont{Baturina, Islamov,
			Bentner, Strunk, Baklanov, and Satta}}]{BaturinaJETP}
	\bibinfo{author}{\bibfnamefont{T.~I.} \bibnamefont{Baturina}},
	\bibinfo{author}{\bibfnamefont{D.~R.} \bibnamefont{Islamov}},
	\bibinfo{author}{\bibfnamefont{J.}~\bibnamefont{Bentner}},
	\bibinfo{author}{\bibfnamefont{C.}~\bibnamefont{Strunk}},
	\bibinfo{author}{\bibfnamefont{M.~R.} \bibnamefont{Baklanov}},
	\bibnamefont{and} \bibinfo{author}{\bibfnamefont{A.}~\bibnamefont{Satta}},
	\bibinfo{journal}{JETP Lett.} \textbf{\bibinfo{volume}{79}},
	\bibinfo{pages}{337} (\bibinfo{year}{2004}).
	
	\bibitem[{\citenamefont{Shahar and Ovadyahu}(1992)}]{Shaharprb}
	\bibinfo{author}{\bibfnamefont{D.}~\bibnamefont{Shahar}} \bibnamefont{and}
	\bibinfo{author}{\bibfnamefont{Z.}~\bibnamefont{Ovadyahu}},
	\bibinfo{journal}{Phys. Rev. B} \textbf{\bibinfo{volume}{46}},
	\bibinfo{pages}{10917} (\bibinfo{year}{1992}),
	\urlprefix\url{http://link.aps.org/doi/10.1103/PhysRevB.46.10917}.
	
	\bibitem[{\citenamefont{Haviland et~al.}(1989)\citenamefont{Haviland, Liu, and
			Goldman}}]{haviprl62}
	\bibinfo{author}{\bibfnamefont{D.~B.} \bibnamefont{Haviland}},
	\bibinfo{author}{\bibfnamefont{Y.}~\bibnamefont{Liu}}, \bibnamefont{and}
	\bibinfo{author}{\bibfnamefont{A.~M.} \bibnamefont{Goldman}},
	\bibinfo{journal}{Phys. Rev. Lett.} \textbf{\bibinfo{volume}{62}},
	\bibinfo{pages}{2180} (\bibinfo{year}{1989}).
	
	\bibitem[{\citenamefont{Parendo et~al.}(2005)\citenamefont{Parendo, Tan,
			Bhattacharya, Eblen-Zayas, Staley, and Goldman}}]{goldmanprl94}
	\bibinfo{author}{\bibfnamefont{K.~A.} \bibnamefont{Parendo}},
	\bibinfo{author}{\bibfnamefont{K.}~\bibnamefont{Tan}},
	\bibinfo{author}{\bibfnamefont{A.}~\bibnamefont{Bhattacharya}},
	\bibinfo{author}{\bibfnamefont{M.}~\bibnamefont{Eblen-Zayas}},
	\bibinfo{author}{\bibfnamefont{N.~E.} \bibnamefont{Staley}},
	\bibnamefont{and} \bibinfo{author}{\bibfnamefont{A.~M.}
		\bibnamefont{Goldman}}, \bibinfo{journal}{Phys. Rev. Lett.}
	\textbf{\bibinfo{volume}{94}}, \bibinfo{pages}{197004}
	(\bibinfo{year}{2005}).
	
	\bibitem[{\citenamefont{Feigel'man et~al.}(2010)\citenamefont{Feigel'man,
			Ioffe, Kravtsov, and Cuevas}}]{FeigAnnals}
	\bibinfo{author}{\bibfnamefont{M.}~\bibnamefont{Feigel'man}},
	\bibinfo{author}{\bibfnamefont{L.}~\bibnamefont{Ioffe}},
	\bibinfo{author}{\bibfnamefont{V.}~\bibnamefont{Kravtsov}}, \bibnamefont{and}
	\bibinfo{author}{\bibfnamefont{E.}~\bibnamefont{Cuevas}},
	\bibinfo{journal}{Annals of Phys.} \textbf{\bibinfo{volume}{325}},
	\bibinfo{pages}{1390} (\bibinfo{year}{2010}).
	
	\bibitem[{\citenamefont{Dubi et~al.}(2007)\citenamefont{Dubi, Meir, and
			Avishai}}]{YonatanNat}
	\bibinfo{author}{\bibfnamefont{Y.}~\bibnamefont{Dubi}},
	\bibinfo{author}{\bibfnamefont{Y.}~\bibnamefont{Meir}}, \bibnamefont{and}
	\bibinfo{author}{\bibfnamefont{Y.}~\bibnamefont{Avishai}},
	\bibinfo{journal}{Nature} \textbf{\bibinfo{volume}{449}},
	\bibinfo{pages}{876} (\bibinfo{year}{2007}).
	
	\bibitem[{\citenamefont{Gantmakher et~al.}(1996)\citenamefont{Gantmakher,
			Golubkov, Lok, and Geim}}]{GantmakherJETP}
	\bibinfo{author}{\bibfnamefont{V.~F.} \bibnamefont{Gantmakher}},
	\bibinfo{author}{\bibfnamefont{M.~V.} \bibnamefont{Golubkov}},
	\bibinfo{author}{\bibfnamefont{J.~G.~S.} \bibnamefont{Lok}},
	\bibnamefont{and} \bibinfo{author}{\bibfnamefont{A.~K.} \bibnamefont{Geim}},
	\bibinfo{journal}{JETP} \textbf{\bibinfo{volume}{82}}, \bibinfo{pages}{951}
	(\bibinfo{year}{1996}).
	
	\bibitem[{\citenamefont{Sambandamurthy
			et~al.}(2005)\citenamefont{Sambandamurthy, Engel, Johansson, Peled, and
			Shahar}}]{murthyprl}
	\bibinfo{author}{\bibfnamefont{G.}~\bibnamefont{Sambandamurthy}},
	\bibinfo{author}{\bibfnamefont{L.~W.} \bibnamefont{Engel}},
	\bibinfo{author}{\bibfnamefont{A.}~\bibnamefont{Johansson}},
	\bibinfo{author}{\bibfnamefont{E.}~\bibnamefont{Peled}}, \bibnamefont{and}
	\bibinfo{author}{\bibfnamefont{D.}~\bibnamefont{Shahar}},
	\bibinfo{journal}{Phys. Rev. Lett.} \textbf{\bibinfo{volume}{94}},
	\bibinfo{pages}{017003} (\bibinfo{year}{2005}).
	
	\bibitem[{\citenamefont{Nguyen et~al.}(2009)\citenamefont{Nguyen, Hollen,
			Stewart, Shainline, Yin, Xu, and Valles}}]{vallesprl103}
	\bibinfo{author}{\bibfnamefont{H.~Q.} \bibnamefont{Nguyen}},
	\bibinfo{author}{\bibfnamefont{S.~M.} \bibnamefont{Hollen}},
	\bibinfo{author}{\bibfnamefont{M.~D.} \bibnamefont{Stewart}},
	\bibinfo{author}{\bibfnamefont{J.}~\bibnamefont{Shainline}},
	\bibinfo{author}{\bibfnamefont{A.}~\bibnamefont{Yin}},
	\bibinfo{author}{\bibfnamefont{J.~M.} \bibnamefont{Xu}}, \bibnamefont{and}
	\bibinfo{author}{\bibfnamefont{J.~M.} \bibnamefont{Valles}},
	\bibinfo{journal}{Phys. Rev. Lett.} \textbf{\bibinfo{volume}{103}},
	\bibinfo{pages}{157001} (\bibinfo{year}{2009}).
	
	\bibitem[{\citenamefont{Sac\'ep\'e et~al.}(2011)\citenamefont{Sac\'ep\'e,
			Dubouchet, Chapelier, Sanque, Ovadia, Shahar, Feigel'man, and
			Ioffe}}]{BenjaminNat}
	\bibinfo{author}{\bibfnamefont{B.}~\bibnamefont{Sac\'ep\'e}},
	\bibinfo{author}{\bibfnamefont{T.}~\bibnamefont{Dubouchet}},
	\bibinfo{author}{\bibfnamefont{C.}~\bibnamefont{Chapelier}},
	\bibinfo{author}{\bibfnamefont{M.}~\bibnamefont{Sanque}},
	\bibinfo{author}{\bibfnamefont{M.}~\bibnamefont{Ovadia}},
	\bibinfo{author}{\bibfnamefont{D.}~\bibnamefont{Shahar}},
	\bibinfo{author}{\bibfnamefont{M.}~\bibnamefont{Feigel'man}},
	\bibnamefont{and} \bibinfo{author}{\bibfnamefont{L.}~\bibnamefont{Ioffe}},
	\bibinfo{journal}{Nat. Phys.} \textbf{\bibinfo{volume}{7}},
	\bibinfo{pages}{239} (\bibinfo{year}{2011}).
	
	\bibitem[{\citenamefont{Sac{\'e}p{\'e}
			et~al.}(2015)\citenamefont{Sac{\'e}p{\'e}, Seidemann, Ovadia, Tamir, Shahar,
			Chapelier, Strunk, and Piot}}]{sacepe2015high}
	\bibinfo{author}{\bibfnamefont{B.}~\bibnamefont{Sac{\'e}p{\'e}}},
	\bibinfo{author}{\bibfnamefont{J.}~\bibnamefont{Seidemann}},
	\bibinfo{author}{\bibfnamefont{M.}~\bibnamefont{Ovadia}},
	\bibinfo{author}{\bibfnamefont{I.}~\bibnamefont{Tamir}},
	\bibinfo{author}{\bibfnamefont{D.}~\bibnamefont{Shahar}},
	\bibinfo{author}{\bibfnamefont{C.}~\bibnamefont{Chapelier}},
	\bibinfo{author}{\bibfnamefont{C.}~\bibnamefont{Strunk}}, \bibnamefont{and}
	\bibinfo{author}{\bibfnamefont{B.}~\bibnamefont{Piot}},
	\bibinfo{journal}{Physical Review B} \textbf{\bibinfo{volume}{91}},
	\bibinfo{pages}{220508} (\bibinfo{year}{2015}).
	
	\bibitem[{\citenamefont{Paalanen et~al.}(1992)\citenamefont{Paalanen, Hebard,
			and Ruel}}]{paalanenprl69}
	\bibinfo{author}{\bibfnamefont{M.~A.} \bibnamefont{Paalanen}},
	\bibinfo{author}{\bibfnamefont{A.~F.} \bibnamefont{Hebard}},
	\bibnamefont{and} \bibinfo{author}{\bibfnamefont{R.~R.} \bibnamefont{Ruel}},
	\bibinfo{journal}{Phys. Rev. Lett.} \textbf{\bibinfo{volume}{69}},
	\bibinfo{pages}{1604} (\bibinfo{year}{1992}).
	
	\bibitem[{\citenamefont{Sambandamurthy
			et~al.}(2004)\citenamefont{Sambandamurthy, Engel, Johansson, and
			Shahar}}]{murthyprl2004}
	\bibinfo{author}{\bibfnamefont{G.}~\bibnamefont{Sambandamurthy}},
	\bibinfo{author}{\bibfnamefont{L.~W.} \bibnamefont{Engel}},
	\bibinfo{author}{\bibfnamefont{A.}~\bibnamefont{Johansson}},
	\bibnamefont{and} \bibinfo{author}{\bibfnamefont{D.}~\bibnamefont{Shahar}},
	\bibinfo{journal}{Phys. Rev. Lett.} \textbf{\bibinfo{volume}{92}},
	\bibinfo{pages}{107005} (\bibinfo{year}{2004}).
	
	\bibitem[{\citenamefont{Sac\'ep\'e et~al.}(2008)\citenamefont{Sac\'ep\'e,
			Chapelier, Baturina, Vinokur, Baklanov, and Sanquer}}]{benjaminprl101}
	\bibinfo{author}{\bibfnamefont{B.}~\bibnamefont{Sac\'ep\'e}},
	\bibinfo{author}{\bibfnamefont{C.}~\bibnamefont{Chapelier}},
	\bibinfo{author}{\bibfnamefont{T.~I.} \bibnamefont{Baturina}},
	\bibinfo{author}{\bibfnamefont{V.~M.} \bibnamefont{Vinokur}},
	\bibinfo{author}{\bibfnamefont{M.~R.} \bibnamefont{Baklanov}},
	\bibnamefont{and} \bibinfo{author}{\bibfnamefont{M.}~\bibnamefont{Sanquer}},
	\bibinfo{journal}{Phys. Rev. Lett.} \textbf{\bibinfo{volume}{101}},
	\bibinfo{pages}{157006} (\bibinfo{year}{2008}).
	
	\bibitem[{\citenamefont{Altshuler et~al.}(2009)\citenamefont{Altshuler,
			Kravtsov, Lerner, and Aleiner}}]{borisprl}
	\bibinfo{author}{\bibfnamefont{B.~L.} \bibnamefont{Altshuler}},
	\bibinfo{author}{\bibfnamefont{V.~E.} \bibnamefont{Kravtsov}},
	\bibinfo{author}{\bibfnamefont{I.~V.} \bibnamefont{Lerner}},
	\bibnamefont{and} \bibinfo{author}{\bibfnamefont{I.~L.}
		\bibnamefont{Aleiner}}, \bibinfo{journal}{Phys. Rev. Lett.}
	\textbf{\bibinfo{volume}{102}}, \bibinfo{pages}{176803}
	(\bibinfo{year}{2009}).
	
	\bibitem[{\citenamefont{Ovadia et~al.}(2009)\citenamefont{Ovadia, Sacepe, and
			Shahar}}]{maozprl}
	\bibinfo{author}{\bibfnamefont{M.}~\bibnamefont{Ovadia}},
	\bibinfo{author}{\bibfnamefont{B.}~\bibnamefont{Sacepe}}, \bibnamefont{and}
	\bibinfo{author}{\bibfnamefont{D.}~\bibnamefont{Shahar}},
	\bibinfo{journal}{Phys. Rev. Lett.} \textbf{\bibinfo{volume}{102}},
	\bibinfo{pages}{176802} (\bibinfo{year}{2009}).
	
	\bibitem[{\citenamefont{Kalok et~al.}(2010)\citenamefont{Kalok, Bilusic,
			Baturina, Vinokur, and Strunk}}]{KalokArxiv}
	\bibinfo{author}{\bibfnamefont{D.}~\bibnamefont{Kalok}},
	\bibinfo{author}{\bibfnamefont{A.}~\bibnamefont{Bilusic}},
	\bibinfo{author}{\bibfnamefont{T.~I.} \bibnamefont{Baturina}},
	\bibinfo{author}{\bibfnamefont{V.~M.} \bibnamefont{Vinokur}},
	\bibnamefont{and} \bibinfo{author}{\bibfnamefont{C.}~\bibnamefont{Strunk}},
	\bibinfo{journal}{arXiv}  (\bibinfo{year}{2010}), \eprint{1004.5153.v1}.
	
	\bibitem[{\citenamefont{Levinson et~al.}(2016)\citenamefont{Levinson, Doron,
			Tamir, Tewari, and Shahar}}]{levinson2016direct}
	\bibinfo{author}{\bibfnamefont{T.}~\bibnamefont{Levinson}},
	\bibinfo{author}{\bibfnamefont{A.}~\bibnamefont{Doron}},
	\bibinfo{author}{\bibfnamefont{I.}~\bibnamefont{Tamir}},
	\bibinfo{author}{\bibfnamefont{G.~C.} \bibnamefont{Tewari}},
	\bibnamefont{and} \bibinfo{author}{\bibfnamefont{D.}~\bibnamefont{Shahar}},
	\bibinfo{journal}{Physical Review B} \textbf{\bibinfo{volume}{94}},
	\bibinfo{pages}{174204} (\bibinfo{year}{2016}).
	
	\bibitem[{Sup()}]{Supplemental}
	\bibinfo{note}{See supplemental material}.
	
	\bibitem[{\citenamefont{Green and Sondhi}(2005)}]{green2005nonlinear}
	\bibinfo{author}{\bibfnamefont{A.~G.} \bibnamefont{Green}} \bibnamefont{and}
	\bibinfo{author}{\bibfnamefont{S.}~\bibnamefont{Sondhi}},
	\bibinfo{journal}{Physical review letters} \textbf{\bibinfo{volume}{95}},
	\bibinfo{pages}{267001} (\bibinfo{year}{2005}).
	
	\bibitem[{\citenamefont{Hogan and Green}(2008)}]{hogan2008universal}
	\bibinfo{author}{\bibfnamefont{P.}~\bibnamefont{Hogan}} \bibnamefont{and}
	\bibinfo{author}{\bibfnamefont{A.}~\bibnamefont{Green}},
	\bibinfo{journal}{Physical Review B} \textbf{\bibinfo{volume}{78}},
	\bibinfo{pages}{195104} (\bibinfo{year}{2008}).
	
	\bibitem[{\citenamefont{Dalidovich and
			Phillips}(2004)}]{dalidovich2004nonlinear}
	\bibinfo{author}{\bibfnamefont{D.}~\bibnamefont{Dalidovich}} \bibnamefont{and}
	\bibinfo{author}{\bibfnamefont{P.}~\bibnamefont{Phillips}},
	\bibinfo{journal}{Physical review letters} \textbf{\bibinfo{volume}{93}},
	\bibinfo{pages}{027004} (\bibinfo{year}{2004}).
	
	\bibitem[{Foo({\natexlab{a}})}]{Footnote1na}
	\bibinfo{note}{For insulators one strives to minimize $I$ in order to avoid
		heating and non-linearities. For technical reasons $I$ is typically in the
		$0.1-1$ nA range.}
	
	\bibitem[{\citenamefont{Doron et~al.}(2016)\citenamefont{Doron, Tamir, Mitra,
			Zeltzer, Ovadia, and Shahar}}]{DoronPRL}
	\bibinfo{author}{\bibfnamefont{A.}~\bibnamefont{Doron}},
	\bibinfo{author}{\bibfnamefont{I.}~\bibnamefont{Tamir}},
	\bibinfo{author}{\bibfnamefont{S.}~\bibnamefont{Mitra}},
	\bibinfo{author}{\bibfnamefont{G.}~\bibnamefont{Zeltzer}},
	\bibinfo{author}{\bibfnamefont{M.}~\bibnamefont{Ovadia}}, \bibnamefont{and}
	\bibinfo{author}{\bibfnamefont{D.}~\bibnamefont{Shahar}},
	\bibinfo{journal}{Phys. Rev. Lett.} \textbf{\bibinfo{volume}{116}},
	\bibinfo{pages}{057001} (\bibinfo{year}{2016}),
	\urlprefix\url{http://link.aps.org/doi/10.1103/PhysRevLett.116.057001}.
	
	\bibitem[{\citenamefont{Garside}(1982)}]{garside1982nucleation}
	\bibinfo{author}{\bibfnamefont{J.}~\bibnamefont{Garside}}, in
	\emph{\bibinfo{booktitle}{Biological mineralization and demineralization}}
	(\bibinfo{publisher}{Springer}, \bibinfo{year}{1982}), pp.
	\bibinfo{pages}{23--35}.
	
	\bibitem[{Foo({\natexlab{b}})}]{FootnoteVDW}
	\bibinfo{note}{In this analogy $V_{escape}$ has a similar role, with an
		opposite $T$-dependence, to the pressure where a first order transition
		occurs in a liquid \cite{DoronPRL}).}
	
	\bibitem[{\citenamefont{Kravtsov}()}]{KravtsovPrivate}
	\bibinfo{author}{\bibfnamefont{V.}~\bibnamefont{Kravtsov}},
	\bibinfo{howpublished}{Private communication}.
	
	\bibitem[{\citenamefont{Kowal and Ovadyahu}(2008)}]{kowal2008scale}
	\bibinfo{author}{\bibfnamefont{D.}~\bibnamefont{Kowal}} \bibnamefont{and}
	\bibinfo{author}{\bibfnamefont{Z.}~\bibnamefont{Ovadyahu}},
	\bibinfo{journal}{Physica C: Superconductivity}
	\textbf{\bibinfo{volume}{468}}, \bibinfo{pages}{322} (\bibinfo{year}{2008}).
	
	\bibitem[{\citenamefont{Kowal and Ovadyahu}(1994)}]{kowal1994disorder}
	\bibinfo{author}{\bibfnamefont{D.}~\bibnamefont{Kowal}} \bibnamefont{and}
	\bibinfo{author}{\bibfnamefont{Z.}~\bibnamefont{Ovadyahu}},
	\bibinfo{journal}{Solid state communications} \textbf{\bibinfo{volume}{90}},
	\bibinfo{pages}{783} (\bibinfo{year}{1994}).
	
	\bibitem[{\citenamefont{Ioffe and Gershenson}(2012)}]{ioffe2012quantum}
	\bibinfo{author}{\bibfnamefont{L.~B.} \bibnamefont{Ioffe}} \bibnamefont{and}
	\bibinfo{author}{\bibfnamefont{M.~E.} \bibnamefont{Gershenson}},
	\bibinfo{journal}{Nature materials} \textbf{\bibinfo{volume}{11}},
	\bibinfo{pages}{567} (\bibinfo{year}{2012}).
	
	\bibitem[{\citenamefont{Parendo et~al.}(2006)\citenamefont{Parendo, Tan, and
			Goldman}}]{parendo2006hot}
	\bibinfo{author}{\bibfnamefont{K.~A.} \bibnamefont{Parendo}},
	\bibinfo{author}{\bibfnamefont{K.~S.~B.} \bibnamefont{Tan}},
	\bibnamefont{and} \bibinfo{author}{\bibfnamefont{A.}~\bibnamefont{Goldman}},
	\bibinfo{journal}{Physical Review B} \textbf{\bibinfo{volume}{74}},
	\bibinfo{pages}{134517} (\bibinfo{year}{2006}).
	
\end{thebibliography}

\pagebreak
\renewcommand{\thefigure}{S\arabic{figure}}
\renewcommand{\thesection}{S\arabic{section}}
\pagenumbering{roman} 
\setcounter{page}{1}
\setcounter{figure}{0}

\large{\bfseries{\centering{Supplemental material\\for\\Instability of insulators near quantum\\phase transitions\\}}}

\section{$\boldsymbol{T_{ph}^{cr}}$ Near $\boldsymbol{B_{c}} $}
\begin{figure}
	\centering
	\includegraphics[width=8.5cm]{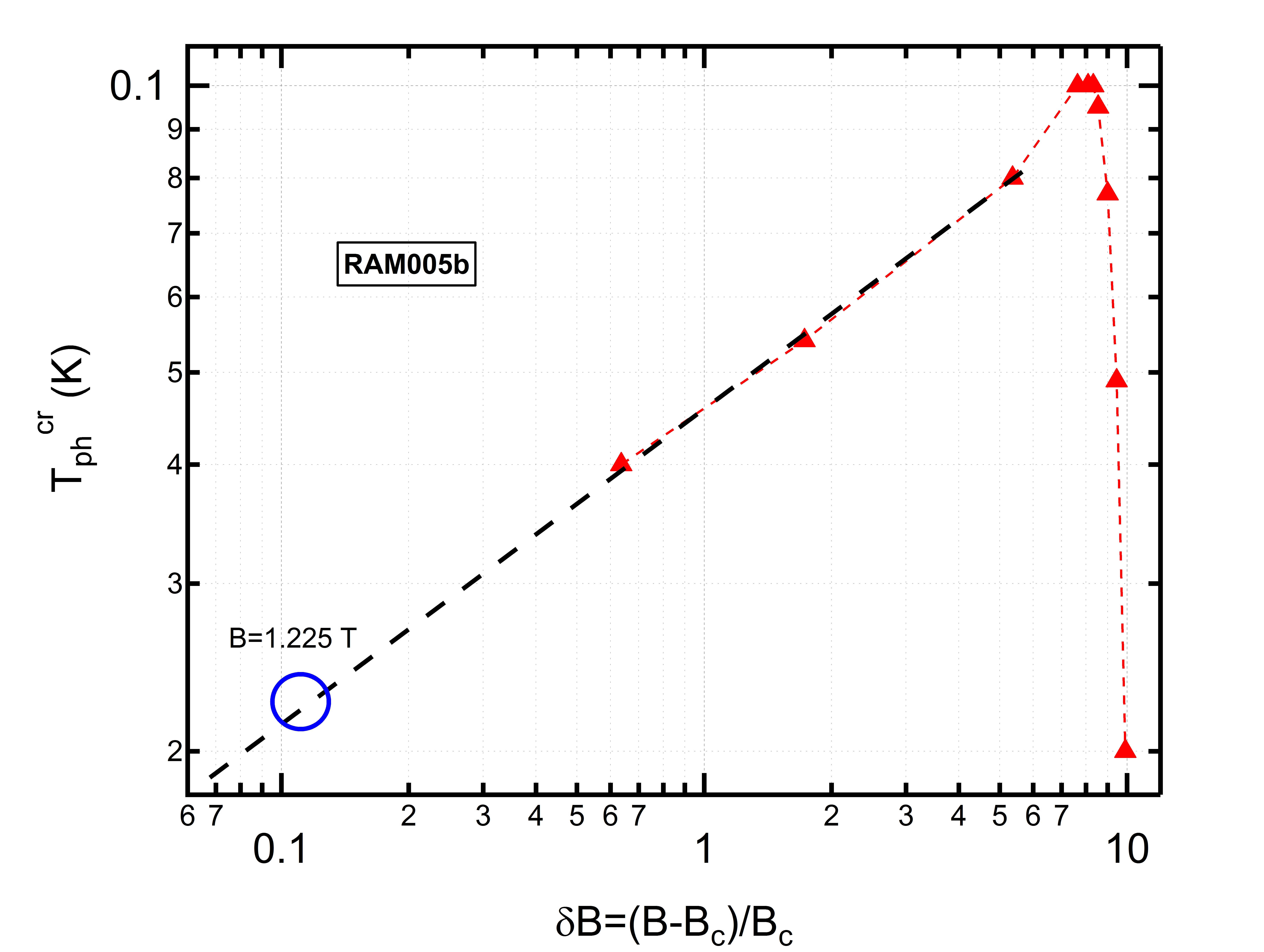}		
	\caption{{\bf $\boldsymbol{T_{ph}^{cr}}$ Near $\boldsymbol{B_{c}} $.} 
		$T_{ph}^{cr}$ vs $\delta B$ (log-log) at various $B$'s. The lowest three data points correspond to $B=1.8, 3$ and $7$ T. The dashed black line marks a power-law according to which at $B=1.225$T $T_{ph}^{cr}\sim 22$ mK (blue circle).
	}			
	\label{FigS1}
\end{figure}

In the inset of figure 2 of the main text we display $I-V$'s measured near $B_{c}$, where below $B=1.225$ T the $I-V$'s seem continuous.
A possible explanation that we raise is that, below this $B$,  the measurement $T=11$ mK might exceed the critical temperature above which there should be no discontinuity ($T_{ph}^{cr}$).
According to ref \cite{borisprl}, $T_{ph}^{cr}\propto \Delta$, where $\Delta$ is the activation $T$ in the insulating phase. 
%$T_{ph}^{cr}\propto \Delta (1+\frac{\beta}{\gamma})^{-\frac{\beta+\gamma}{\beta \gamma}}$
In addition, approaching the quantum critical point, all energy scales should go to zero, therefore $\Delta \to 0$ as $\delta B\equiv \frac{B-B_{C}}{B_{C}}\to 0$ \cite{sondhirmp}.
Therefore, if we assume that sufficiently close to $B_{c}$, $\Delta \propto \delta B^{\lambda}$, where $\lambda$ is a constant, we get that $T_{ph}^{cr}\propto \delta B^{\lambda}$.
In figure \ref{FigS1} we display $T_{ph}^{cr}$ vs $\delta B$ (log-log) at various $B$'s. 
The dashed black line is a power-law fit for the lowest three data points ($B=1.8, 3$ and $7$ T respectively). 
Extrapolating it to $B=1.225$ T, we get that $T_{ph}^{cr}\sim 22$ mK, higher than the measurement $T=11$ mK.
Although this analysis is not rigorous, it does imply that a decrease in $T_{ph}^{cr}$ is not the cause of the continuous $I-V$'s below $B=1.225$.

\section{Contacts}
The contacts of sample RAM005b were prepared by pressed indium on a Au wire. 
While Indium is a type 1 superconductor the data presented in the manuscript was measured at $B$'s higher than $H_{c}$ of Indium ($H_{c}\sim$0.028 T, while the data is of $B>0.5$ T) therefore during these measurements, the contacts were in a normal state. 
The contacts of sample BT1c are Ti/Au pads (15nm Au over 10nm Ti), patterned via optical lithography and prepared prior to the deposition of the In$_2$O$_3$. 
In order to verify that contact resistance is negligible we compared 4-terminal and 2-terminal measurements at various $B$'s and in a $R$ range where both measurements are expected to yield similar results ($R<1$M$\Omega$) and found that the discrepancies were negligible.

\section{Extracting the measured 4-probe results from the low resistive state of the $\boldsymbol{I-V}$'$\boldsymbol{s}$}
\begin{figure}
	\centering
	\includegraphics[width=8.5cm]{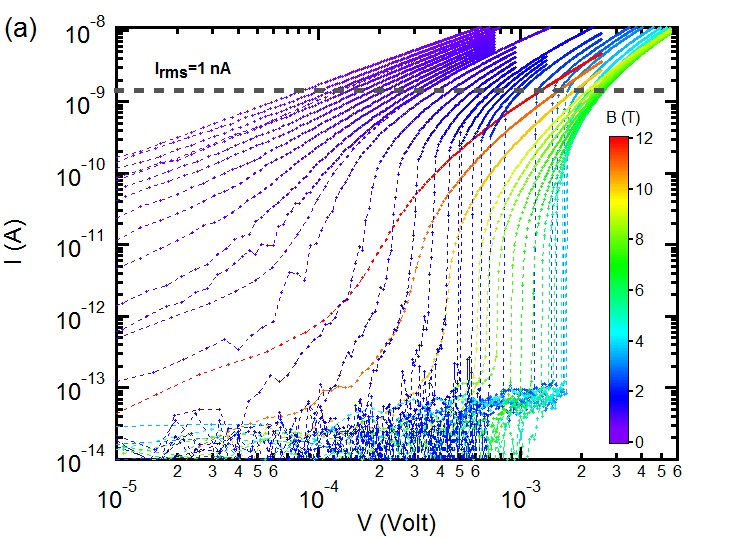}		
	\includegraphics[width=8.5cm]{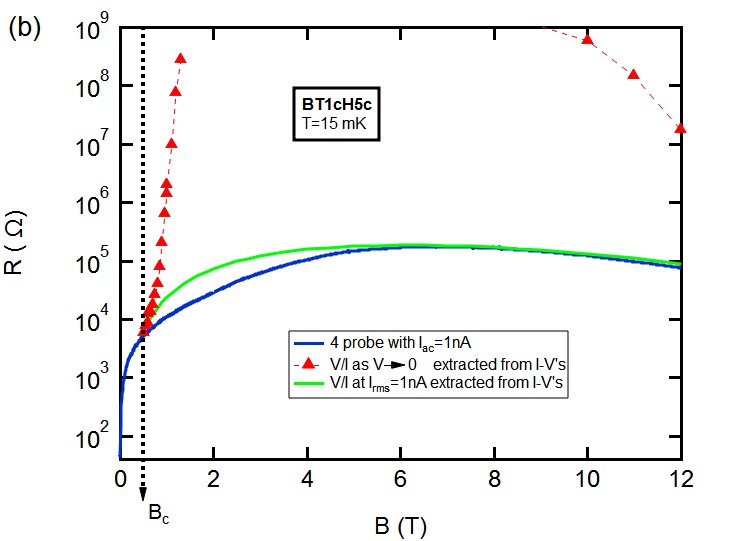}
	\caption{{\bf Extracting the 4-terminal results of figure 5 (main text) from the LR state.} 
		(a) $I-V$'s of sample BT1c, measured at $T=15$mK at various $B$'s. The black dashed line correspond to $I_{rms}=1$nA. 
		(b) $R$ (log scale) vs. $B$ of sample BT1c at $T=15$ mK. The red triangles and blue line also appear in figure 5 of the main text and correspond, respectively, to $R$ extracted from the $I-V$'s in the limit $V \to 0$ and a 4-probe AC measurement with $I_{rms}=1$ nA. The green line is extracted from the same $I-V$'s as the red data but at $I=\sqrt{2}*1$nA. 
	}			
	\label{FigS2}
\end{figure}

In figure 5 of the main text we display a large discrepancy between a 2-terminal $V \to 0$ measurement and a 4-terminal, $I$-biased, measurement where we used a standard current of $I_{rms}=1$nA.
Here we show that the low $R$ measured by 4-terminal is a results of heating where the injected current drives the sample to the low resistive (LR) state.
Displayed in figure \ref{FigS2}a are the $I-V$'s of sample BT1c, measured at $T=15$mK at various $B$'s. 
The dashed black line marks $I_{rms}=1$nA. 
It can be seen that at $B$'s that exhibit a discontinuity this 1nA line crosses the LR state.
In order to estimate the "$R$" of a $I$ biased measurement of $I_{rms}=1$nA from these dc $I-V$'s we used the simplest approximation of $R=V/I$ at $I=1nA$. The result of this analysis is displayed as the green line in figure \ref{FigS2}b, where the red and blue data are the same as in figure 5 of the main text.
It can be seen that although this rough approximation does not exactly reproduce the 4-terminal result, it does predict similar $R$'s.

\section{Unavoidable "transport catastrophe" in a larger sample}	
\begin{figure}[h!]
	\centering
	\includegraphics[width=8.5cm]{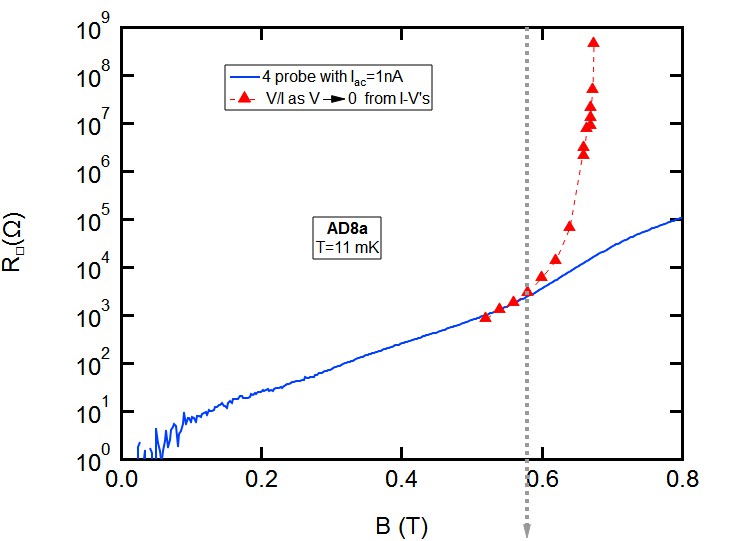}		
	\caption{{\bf Unavoidable "transport catastrophe".} 
		$R$ (log scale) vs. $B$ of sample AD8a at $T=11$ mK. The red triangles were extracted from the $I-V$'s in the limit $V \to 0$. The blue line was measured using a standard 4-probe AC measurement with $I_{rms}=1$ nA. 
	}			
	\label{FigS3}
\end{figure}

The cross section of sample BT1c (of figure 5 of the main text) is 100 times smaller than that of sample RAM005b (figures 1-4 of the main text), therefore passing the same $I$ in both samples results in a 100 times larger current density, J, in BT1c than in RAM005b. To show that the large discrepancy, displayed in figure 5 of the main text and figure \ref{FigS2}b, is not due to the sample's size we repeated these measurements on sample AD8a, which is of size 2$\times$1 mm and 30 nm thick (therefore it's cross section is twice as large as of sample RAM005b). In figure \ref{FigS3} we display the comparison between the 2-probe $R$ extracted from the linear part of the $I-V$'s (red triangles) and 4-probe $R$ using an excitation current of 1nA. The results are very similar to those displayed in figure 5 of the main text.	

\section{2-terminal and 4-terminal measurement configurations}
\begin{figure}
	\centering
	\includegraphics[width=8.5cm]{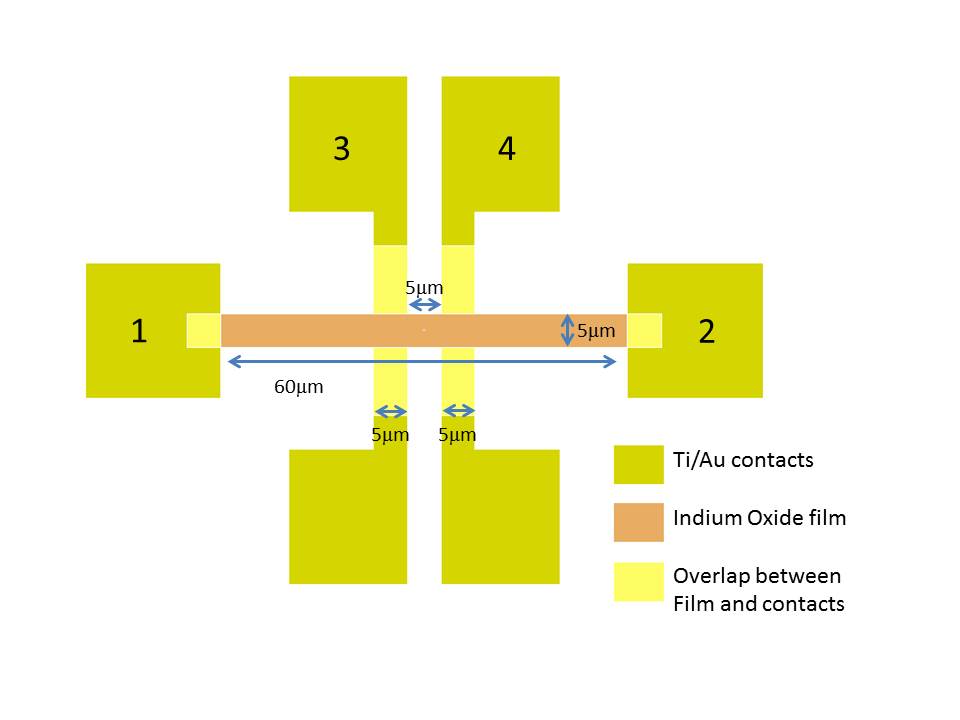}		
	\caption{{\bf Sketch of sample BT1c.} 
		Sample BT1c is Hall-bar shaped. The dark yellow sections mark the Ti/Au contacts, the orange section marks the thin film of amorphous Indium-Oxide and the light yellow marks the overlap between the film and the contacts. The relevant lengths appear on the sketch. 
	}			
	\label{FigS4}
\end{figure}

In figure \ref{FigS4} we display a sketch of sample BT1c. 
4-terminal measurement were performed by passing an AC current between the far contacts (numbers 1 and 2 in the figure) while measuring the voltage drop across the center contacts (3 and 4).
The 2 terminal measurements were performed by varying a DC bias $V$ between the far contacts (1 and 2) and measuring the resulting $I$ through the same contacts. 
The V/I data displayed in figure 5 of the main text and figure \ref{FigS2}b were normalized by the number of squares. 
The reason we performed the 2 terminal measurements using the far contacts is to make sure that in both 4 and 2 terminal measurements the $I$ flows in the sample through the same path.

\section{$\boldsymbol{I-V}$'s near the metal-insulator transition}
\begin{figure} [h!]
	\centering
	\includegraphics[width=8.5cm]{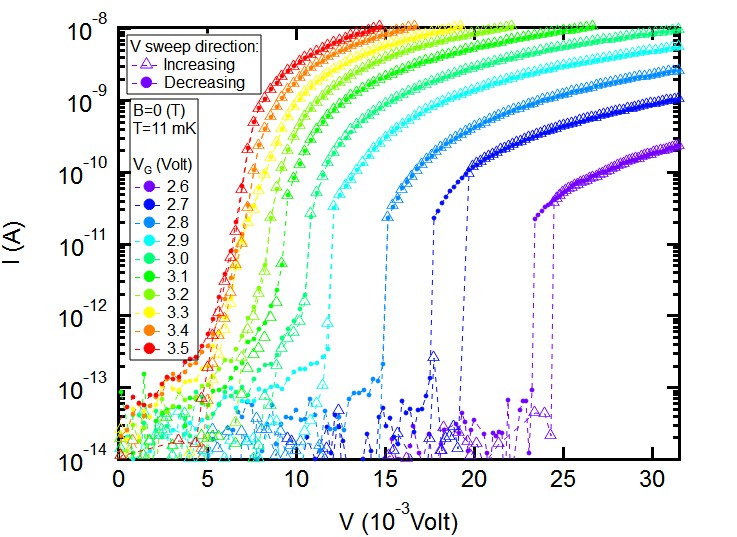}		
	\caption{{\bf $\boldsymbol{I-V}$'s of Silicon MOSFET near the metal-insulator transition.} 
		$I$ (log-scale) vs $V$ measured on a silicon MOSFET sample at $T=11$ mK and $B$=0. Triangles (circles) mark an increasing (decreasing) $V$ sweep and the color-coding stands for the applied $V_{G}$. It can be seen that the $I-V$'s show large $I$ discontinuities where increasing $V_{G}$ results in a decrease $V_{escape}$.
	}			
	\label{FigS5}
\end{figure}
In the discussion section of the main-text we claim that the insulating instability is general and is also relevant to other insulators near quantum critical points. 
To verify this claim we measured a Silicon metal-oxide-semiconductor field-effect transistor (Si-MOSFET) in the insulating phase of the gate voltage ($V_{G}$) induced metal-insulator transition (MIT). 
In figure \ref{FigS5} we display the $I-V$'s of this Si-MOSFET at $T=11$ mK and $B=0$ where different colors mark different values of $V_{G}$. 
The $I-V$'s are remarkably similar to those measured in the insulating phase near the SIT in the sense that there are large current discontinuities and  that approaching the MIT these discontinuities appear at lower $V$'s.
A full characterization of the $I-V$ discontinuities near the MIT will appear in a future publication.

\end{document}